\documentclass[twocolumn]{amsart}

\usepackage{amssymb,amsmath,amscd,graphicx}

\newcommand{\ts}{\hspace{0.5pt}}
\newcommand{\nts}{\hspace{-0.5pt}}
\newcommand{\RR}{\mathbb{R}\ts}
\newcommand{\ZZ}{\mathbb{Z}}
\newcommand{\NN}{\mathbb{N}}
\newcommand{\Nnull}{\mathbb{N}_{0}^{}}
\newcommand{\QQ}{\mathbb{Q}}

\newcommand{\cL}{\mathcal{L}}
\newcommand{\cO}{\mathcal{O}}
\newcommand{\vL}{\varLambda}
\newcommand{\vS}{\varSigma}

\newcommand{\dd}{\,\mathrm{d}}
\newcommand{\ee}{\ts\mathrm{e}}
\newcommand{\ii}{\mathrm{i}\ts}

\newcommand{\myfrac}[2]{\frac{\raisebox{-2pt}{$#1$}}
      {\raisebox{0.5pt}{$#2$}}}

\newcommand{\sB}{\ts\underline{\nts B\!}\,}
\newcommand{\defeq}{\mathrel{\mathop:}=}
\newcommand{\eqdef}{=\mathrel{\mathop:}}

\DeclareFontFamily{U}{mathx}{\hyphenchar\font45}
\DeclareFontShape{U}{mathx}{m}{n}{ <5> <6> <7> <8> <9> <10>
   <10.95> <12> <14.4> <17.28> <20.74> <24.88> mathx10 }{}
\DeclareSymbolFont{mathx}{U}{mathx}{m}{n}
\DeclareFontSubstitution{U}{mathx}{m}{n}
\DeclareMathAccent{\widecheck}{0}{mathx}{"71}

\DeclareMathOperator{\dens}{dens}
\DeclareMathOperator{\sinc}{sinc}
\DeclareMathOperator{\vol}{vol}

\begin{document}

\twocolumn[
\begin{LARGE}
\centerline{Inflation versus projection sets in aperiodic systems:}
\centerline{The role of the window in averaging and diffraction}\vspace{3ex}
\end{LARGE}
\centerline{\large Michael Baake$^{1,3}$ and Uwe Grimm$^{2,3}$}\vspace{2ex}
\begin{footnotesize}
\centerline{
${}^{1}$\textit{Fakult\"at f\"ur Mathematik, Universit\"at Bielefeld,
  Postfach 100131, 33501 Bielefeld, Germany\vspace{1mm}}}
\centerline{${}^{2}$\textit{School of Mathematics and Statistics, 
The Open University, Walton Hall, Milton Keynes MK7 6AA, UK\vspace{1mm}}}
\centerline{${}^{3}$\textit{School of Natural Sciences, 
University of Tasmania, Private Bag 37, 
Hobart TAS 7001, Australia\vspace{1mm}}}
\end{footnotesize}\vspace{4ex}
\begin{small}
\hrule\vspace{2ex}
\begin{minipage}{\textwidth}
  \textbf{Abstract}\vspace{2ex}\\ Tilings based on the cut and project
  method are key model systems for the description of aperiodic
  solids. Typically, quantities of interest in crystallography involve
  averaging over large patches, and are well defined only in the
  infinite-volume limit. In particular, this is the case for
  autocorrelation and diffraction measures. For cut and project
  systems, the averaging can conveniently be transferred to internal
  space, which means dealing with the corresponding windows. In this
  topical review, we illustrate this by the example of averaged
  shelling numbers for the Fibonacci tiling and recapitulate the
  standard approach to the diffraction for this example. Further, we
  discuss recent developments for cut and project structures with an
  inflation symmetry, which are based on an internal counterpart of
  the renormalisation cocycle. Finally, we briefly review the notion
  of hyperuniformity, which has recently gained popularity, and its
  application to aperiodic structures.
  \vspace{2ex}
\end{minipage}\vspace{2ex}
\hrule
\end{small}\vspace{6ex}
]

\section{Introduction}

The discovery of quasicrystals in the early 1980s \cite{SBGC} not only
led to a reconsideration of the fundamental concept of a crystal (see
\cite{G15} and references therein), but also highlighted the need for
a mathematically robust treatment of the diffraction of systems that
exhibit aperiodic order. The foundations for a rigorous approach were
laid by Hof \cite{Hof1}. In particular, the measure-theoretic approach
via the autocorrelation and diffraction measures allows for a
mathematically rigorous discussion and separation of the different
spectral components, the pure point, singular continuous and
absolutely continuous parts; see \cite{BG12} for background and
examples, and \cite[Sec.~9]{TAO} for a systematic exposition. For
general background on the theory of aperiodic order, we refer to
\cite{PF,AS,Q,TAO,KLS,Morlet18} and references therein.

Within a few years, it was established that regular model sets
\cite{Moody00} (systems obtained by projection from higher-dimensional
lattices via cut and project mechanisms with `nice' windows) have pure
point diffraction \cite{Martin,RS17a}. We refer to the discussion in
\cite{TAO} for details and examples, and to \cite{BEG} for an
instructive application of the cut and project approach to an
experimentally observed structure with twelvefold symmetry.  The
result on the pure point nature of diffraction holds for rather
general setups, including cut and project schemes with non-Eucliden
internal spaces. It has recently been generalised to weak model sets
of extremal densities \cite{BHS,RS17b}, for which the window may even
entirely consist of boundary, that is, has no interior; see also
\cite{Str17,Str20} for recent work on pure point spectra.

While systems based on a cut and project scheme are generally well
understood, this is less so for systems originating from substitution
or inflation rules, which constitute another popular method of
generating systems with aperiodic order; see \cite{Q,TAO,Dirk} and
references therein for details. There has been recent progress
particularly on substitutions of constant length; see
\cite{Neil,Bart,BS19,BCM,BGM19,BS20}.

There are familiar examples of inflation-based structures for all
spectral types, such as the Fibonacci chain for a pure point
diffractive system, the Thue--Morse chain for a system with purely
singular continuous diffraction, and the binary Rudin--Shapiro chain
as the paradigm of a system with absolutely continuous diffraction;
see \cite{PF,AS,TAO} for details.  When one equips the Rudin--Shapiro
chain with balanced weights ($\pm 1$), it becomes homometric with the
binary Bernoulli chain with random weights $\pm 1$ \cite{BG09}. It is
easy to construct inflation-based systems which combine any of these
spectral components in their diffraction; see \cite{BGG} for
examples. As of today, the celebrated Pisot substitution conjecture
(which stipulates that an irreducible Pisot substitution has pure
point spectrum) remains open; see \cite{Aki} for a review of the state
of affairs.

While diffraction was the first property to be analysed in detail,
many other questions from traditional crystallography and lattice
theory require an extension to their aperiodic counterparts
\cite{BZ}. In particular, classic counting problems based on lattices,
when reformulated for point sets in aperiodic tilings, need both a
conceptual reformulation and new tools to tackle them. The key
observation is the necessity to employ averaging concepts, and then
tools from dynamical systems and ergodic theory \cite{Q,Sol97,TAO}. If
one is in the favourable situation of point sets that emerge from
either the projection formalism or an inflation procedure, many of
these averaged quantities are well defined and can actually be
calculated; see \cite{BG03} and references therein.  Despite good
progress, many questions in this context remain open.\smallskip

Let us sketch how this introductory review is organised.  Our guiding
example in this exposition is the classic, self-similar Fibonacci
tiling of the real line. Its descriptions as an inflation set and as a
cut and project set are reviewed in Section~\ref{sec:Fibo}. As a
simple example of the role of the window in averaging, we discuss the
averaged shelling for this system in Section~\ref{sec:shell}. This is
followed by a brief review of the standard approach to diffraction in
Section~\ref{sec:standard}, where we exploit the description of the
Fibonacci point set as a cut and project set and the general results
for the diffraction of regular model sets.

In Section~\ref{sec:cocycle}, we recapitulate the recently developed
internal cocycle approach. For systems which possess both an inflation
and a projection interpretation, such as the Fibonacci tiling, the
inflation cocycle can be lifted to internal space. This makes it
possible to efficiently compute the diffraction of certain cut and
project systems with complicated windows, such as windows with fractal
boundaries, as are commonly found in inflation structures. To explore
this further, we reconsider planar examples, based on the Fibonacci
substitution, in Section~\ref{sec:planar}.

Finally, in Section~\ref{sec:hyper}, we discuss the use of
`hyperuniformity' as a measure of order in Fibonacci systems. This
amounts to an investigation of the asymptotic behaviour of the total
diffraction intensity near the origin. It turns out that this can
dinstinguish between generic and inflation-invariant choices for the
window in the cut and project scheme.

\section{The Fibonacci tiling revisited}\label{sec:Fibo}

Let us start with a paradigm of aperiodic order in one dimension, the
classic Fibonacci tiling. It can be defined via the primitive
two-letter inflation rule
\[
    \varrho \colon \quad a\mapsto ab\ts , \quad b \mapsto a\ts ,
\]
where $a$ and $b$ represent \emph{tiles} (or intervals) of length
$\tau = \frac{1}{2} \bigl( 1 + \mbox{\small $\sqrt{5}$} \, \bigr)$ and
$1$, respectively. The corresponding incidence matrix is given by
\begin{equation}\label{eq:fibmat}
    M \, = \, \begin{pmatrix} 1 & 1 \\ 1 & 0 \end{pmatrix},
\end{equation}
which has Perron--Frobenius eigenvalue $\tau$. Its left and right
eigenvectors read
\begin{equation}\label{eq:ev}
  \langle u\ts | \, = \, \myfrac{\tau+2}{5} \bigl(\tau,1\bigr)\quad
\text{and} \quad
   |\ts v\rangle \, = \, \bigl(\tau^{-1},\tau^{-2}\bigr)^{T},
\end{equation}
where we employ Dirac's intuitive `bra-c-ket' notation, which makes it
easy to distinguish row and column vectors.  We normalise the right
eigenvector $|\ts v\rangle$ such that $\langle 1 | \ts v \rangle =1$,
which means that its entries are the \emph{relative frequencies} of
the tiles. For later convenience, we normalise the left eigenvector
$\langle u\ts |$ by setting $\langle u\ts |\ts v\rangle=1$, rather
than using the vector of natural tile lengths itself. With this
normalisation, we have
\begin{equation}\label{eq:Pdef}
\begin{split}
    \lim_{n\to\infty} \tau^{-n} M^n \, &= \, \myfrac{\tau+2}{5}
    \begin{pmatrix} 1 & \tau^{-1} \\ 
                   \tau^{-1} & \tau^{-2}\end{pmatrix}\\[1mm]
     &= \, |\ts v\rangle\langle u\ts | \, \eqdef \, P\ts ,
\end{split}
\end{equation}
where $P=P^2$ is a symmetric projector of rank $1$ with spectrum
$\{1,0\}$.

Starting from the legal seed $b|a$, where the vertical bar denotes 
the origin, and iterating the square of the inflation rule $\varrho$
generates a tiling of the real line that is invariant under
$\varrho^2$; see \cite[Ex.~4.6]{TAO} for details and why it does not
matter which of the two fixed points of $\varrho^2$ one chooses. Let
us use the left endpoints of each interval as \emph{control points}
and denote the set of these points by $\vL_{a}$ and $\vL_{b}$,
respectively. Clearly, since $0\in\vL_{a}$ and all tiles have either
length $\tau$ or length $1$, all coordinates are integer linear
combinations of these two tile lengths, and we have
\[
   \vL_{a,b}\, \subset\, \ZZ[\tau]\, =\, \{m+n\tau : m,n\in\ZZ\}\ts . 
\]
The incidence matrix $M$ only contains information about the number of
tiles under inflation, but not about their positions. To capture the
latter, and thus encode the full information of the inflation, we
consider the set-valued \emph{displacement matrix}
\begin{equation}\label{eq:T}
    T \, = \, \begin{pmatrix} \{ 0 \} & \{ 0 \} \\
    \{\tau\} & \varnothing \end{pmatrix},
\end{equation}
where $\varnothing$ denotes the empty set. Note that $T$ is the
geometric counterpart of the instruction matrices that are used in the
symbolic context \cite{Q}. The matrix elements of $T$ are sets that
specify the relative displacement for all tiles under inflation. For
instance, the two entries in the first column correspond to a long
tile with relative shift $0$ and a small tile with shift $\tau$
originating from inflating a long tile. Clearly, the inflation matrix
$M$ is recovered if one takes the elementwise cardinality of $T$,
noting that the empty set has cardinality $0$.

The inflation rule $\varrho$ induces an iteration on pairs of 
point sets, namely  
\begin{equation}\label{eq:vL-rec}
\begin{split}
   \vL^{(n+1)}_{a} \, &= \, \tau\vL^{(n)}_{a}\cup 
   \tau\vL^{(n)}_{b}\ts ,\\
   \vL^{(n+1)}_{b}\, &= \, \tau\vL^{(n)}_{a}\nts +\tau\ts ,
\end{split}
\end{equation}
with suitable initial conditions $\vL^{(0)}_{a,b}$.  When one starts
with the left endpoints of a legal seed, this iteration precisely
reproduces the endpoints of the corresponding, successive inflation
steps.  In this case, the union on the right-hand side is disjoint. In
particular, for the above choice of $\vL_{a,b}$, one needs
$\vL^{(0)}_{a}=\{0\}$ and $\vL^{(0)}_{b}=\{-1\}$.

The point sets $\vL_{a,b}$ also have an interpretation as a cut and
project set. Here, we use the natural (Minkowski) embedding of the
module $\ZZ[\tau]$ in the plane $\RR^{2}$, by associating to each
$x=m+n\tau\in\ZZ[\tau]$ its image
$x^{\star}=m+n\tau^{\star}=m+n(1-\tau)$ under algebraic conjugation
(which maps $\sqrt{5}$ to $-\sqrt{5}\,$). This gives
\begin{align*}
   \cL \, &= \, \bigl\{(x,x^{\star}) : x\in\ZZ[\tau]\bigr\}\\
    & = \, \bigl\{(m+n\tau,m+n\tau^{\star}) : m,n\in\ZZ\bigr\}\\
    & = \, \bigl\{m(1,1) + n(\tau,\tau^{\star}): m,n\in\ZZ\bigr\},
\end{align*}
which is a planar lattice with basis vectors $(1,1)$ and
$(\tau,\tau^{\star})$; see \cite{TAO,BEG} for details and further
examples. Here, we refer to the two one-dimensional subspaces 
of $\RR^2=\RR\times\RR$ as the \emph{physical} and the
\emph{internal} space, respectively. The physical space hosts our
point sets $\vL_{a,b}$, while the windows are subsets of the internal
space, with the \mbox{$\star$-map} providing the relevant link 
between the two spaces.

\begin{figure}
\includegraphics[width=\columnwidth]{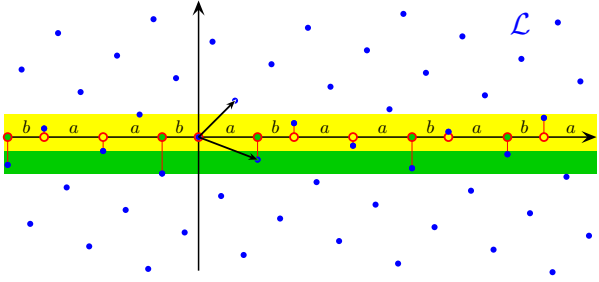}
\caption{Cut and project description of the Fibonacci chain from the
  lattice $\cL$ (blue dots). The windows $W_{a}$ and $W_{b}$ are the
  cross-sections of the yellow and green strips,
  repectively.\label{fig:fiboproj}}
\end{figure}

The two point sets $\vL_{a,b}$ are given by the projection of all
points of $\cL$ within two strips; compare
Figure~\ref{fig:fiboproj}. These strips are defined by their
cross-sections, usually called \emph{windows}, which are the half-open
intervals $W_{a}=[\tau-2,\tau-1)$ and $W_{b}=[-1,\tau-2)$.  With
$L = \ZZ[\tau]$, the projection of $\cL$ into physical space, the
point sets are thus given by
\begin{equation}\label{eq:vL-ab}
   \vL_{a,b} \, = \, \bigl\{x\in L : 
   x^{\star}\in W_{a,b}\bigr\}\ts .
\end{equation}

One of the powerful properties of the cut and project approach is that
we can switch between the physical space and the internal space, and
calculate properties in the latter.  Taking the $\star$-image of
\eqref{eq:vL-rec}, we obtain the relations
\begin{equation}\label{eq:W-rec}
    W_{a} \, = \, \sigma W_{a} \cup \sigma W_{b}\ts ,
    \qquad W_{b} \, = \, \sigma W_{a} + \sigma\ts ,
\end{equation}
where $\sigma=\tau^{\star}=1-\tau$ satisfies
$\lvert\sigma\rvert<1$. These relations are an important ingredient
for the internal cocycle approach. Due to $\lvert\sigma\rvert<1$, this
gives rise to a contractive iterated function system, which has the
windows $W_{a,b}$ (or, more precisely, their closures) as its unique
solution.

One key property, which can be employed to show that the point sets
$\vL_{a,b}$ are pure point diffractive, is the fact that the
$\star$-images of $\vL_{a,b}$ are \emph{uniformly distributed} in the
windows $W_{a,b}$, which makes it possible to translate the
computation of \emph{averaged quantities} in physical space to
computations in internal space.

\section{Shelling}\label{sec:shell}

Let us discuss a simple example of an averaged quantity, the averaged
shelling function for the Fibonacci point set; see \cite{BG03} for the
concept and various applications to aperiodic systems.  The shelling
problem is related to the autocorrelation as well as to diffraction;
we include it here to demonstrate, in a simple explicit example, the
advantages of using internal space for this type of analysis.

For a point set, the \emph{shelling} problem asks for the number
$n(r,x)$ of points that lie on shells of radius $r$, taken with
respect to a fixed centre $x$. For an aperiodic point set, this
generally depends on the choice of the centre. The \emph{averaged
  shelling} numbers $a(r)$ are obtained by taking the average over all
choices of centres, where we limit ourselves to centres that are
themselves in the point set, so $x\in\vL$. Clearly, since we are
dealing with a one-dimensional point set, any shell can have at most
two points, so $n(r,x)\in\{0,1,2\}$ for all $r\in\RR$, with $n(r,x)=0$
if $r\not\in \ZZ[\tau]$, as well as $n(0,x)=a(0)=1$. Clearly, this
also implies that $a(r)\in [0,2]$ for all $r\in\RR$, with $a(r)=0$
whenever $r\not\in \ZZ[\tau]$.

Consider a point $x\in\vL$ and $r=m+n\tau\in\ZZ[\tau]$. To compute
$n(r,x)$, we have to check whether $x\pm r$ are also in the point set
$\vL$. From the model set description, we know that $x^{\star}\in
W\!$, and checking whether $x\pm r$ are in $\vL$ is equivalent to
checking whether $x^{\star}\pm r^{\star}\in W\!$. In other words, we
can express $n(r,x)$ for $r>0$ in terms of the window $W\!$ as
\[
    n(r,x) \, = \, 1^{}_{W}(x^{\star})1^{}_{W}(x^{\star}\! +r^{\star}) + 
    1^{}_{W}(x^{\star})1^{}_{W}(x^{\star}\! -r^{\star})\ts ,
\]
where $1^{}_{W}$ denotes the indicator (or characteristic) 
function of the window $W\!$, defined by
\[
   1^{}_{W}(x)\, = \, \begin{cases} 1, & \text{if $x\in W\!$,}\\
   0, & \text{otherwise.}\end{cases}
\]
While it is possible to perform this computation for any given value
of $x$ and $r$, there is no simple closed formula for these
coefficients.

\begin{figure}
\includegraphics[width=0.49\columnwidth]{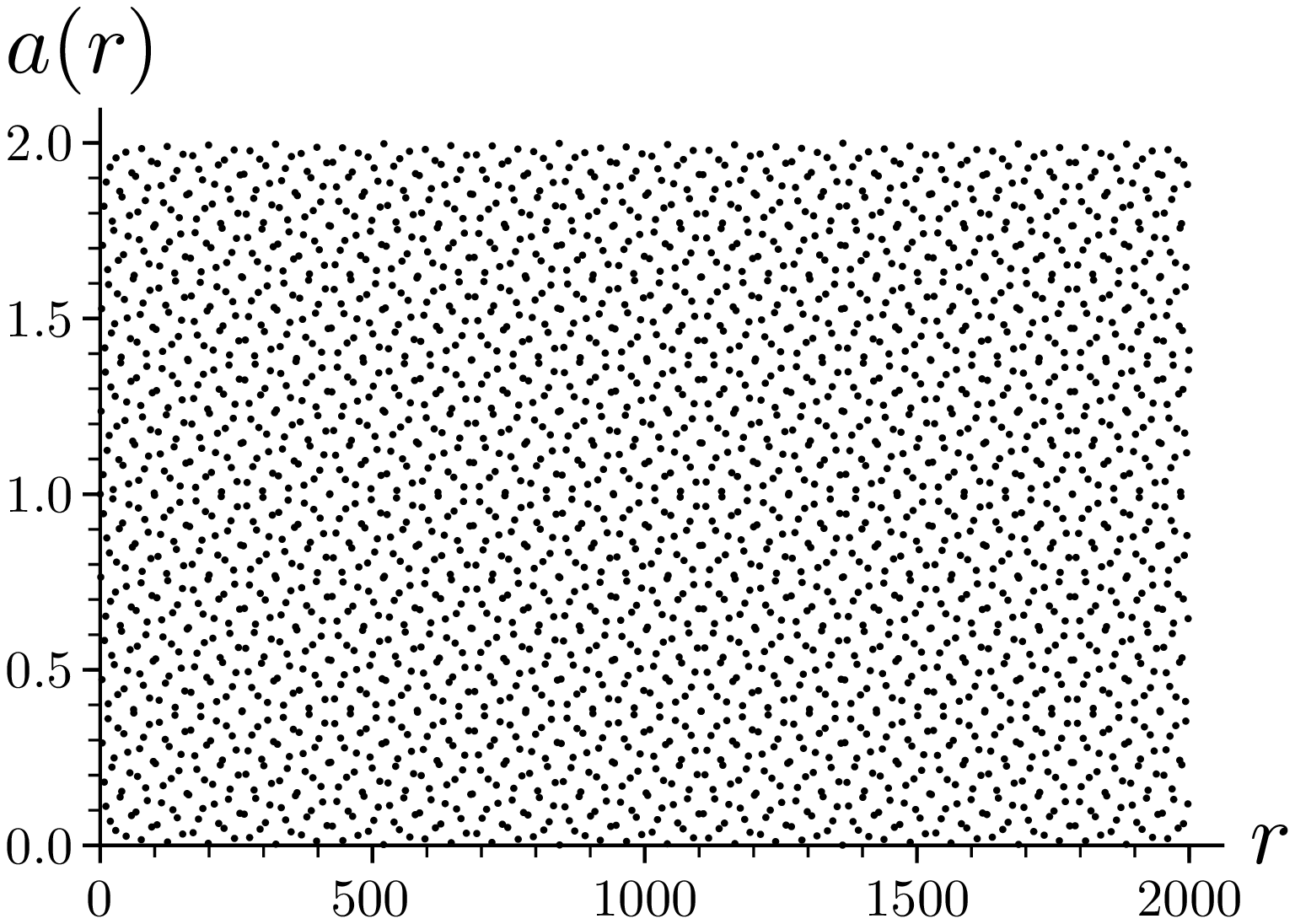}\hfill
\includegraphics[width=0.49\columnwidth]{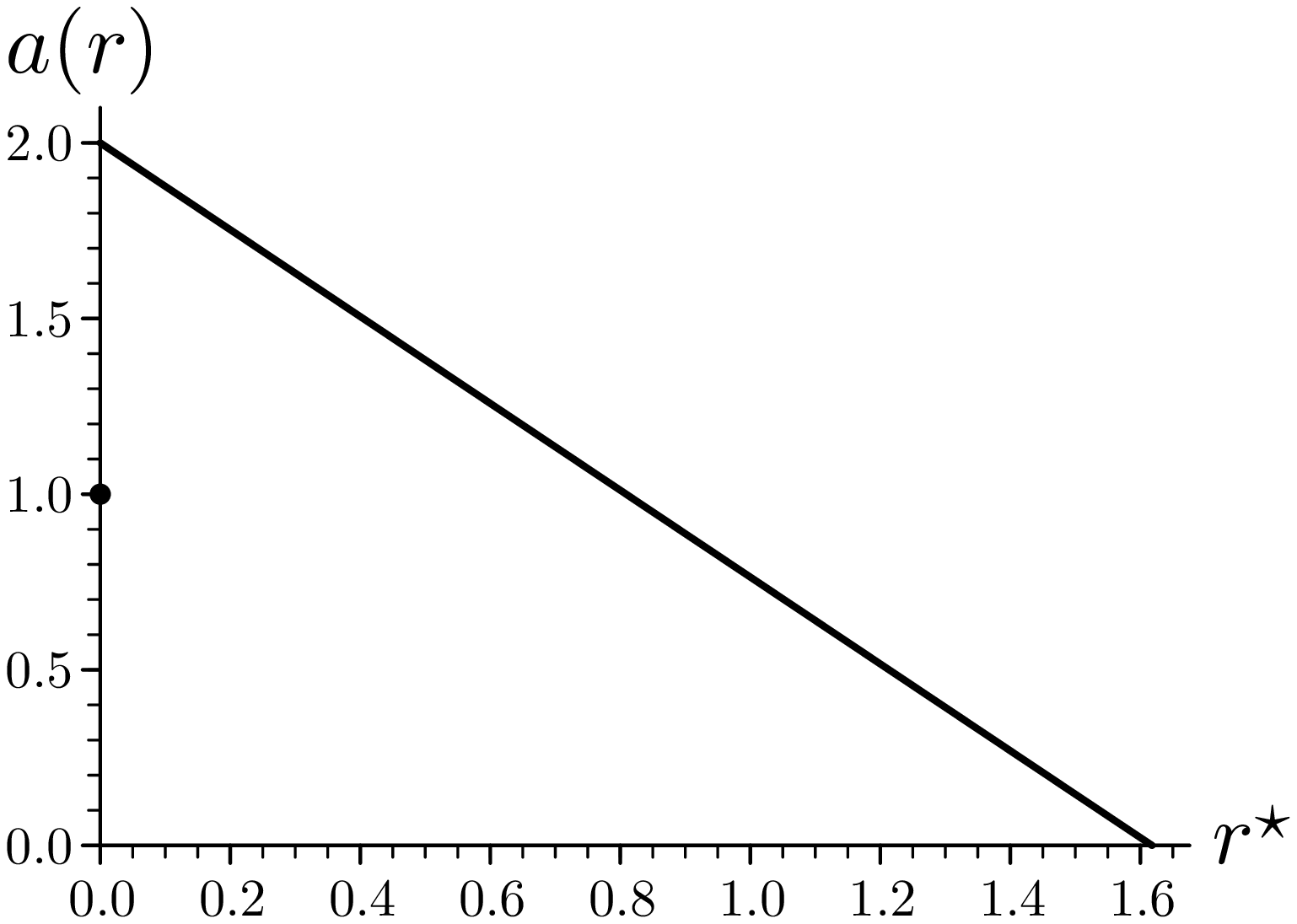}
\caption{Averaged shelling numbers $a(r)$ for the Fibonacci point set
  as a function of $r$ (left) and $r^{\star}$
  (right).\label{fig:fiboshell}}
\end{figure}

To obtain the averaged shelling number, we have to consider all 
$x\in\vL$ as centres, each with the same weight, which means 
averaging over all $x^{\star}\in W\!$. Define $\nu(r)=\nu(-r)$ 
as the relative frequency to find one point of $\vL$ at $x$ as 
well as one at $x+r$, so $a(0)=\nu(0)=1$ and $a(r)=2\nu(r)$ for
$r>0$, to account for the points on both sides. Now, for
$r\in\ZZ[\tau]$, the frequency $\nu(r)$ of having both $x^{\star}\in
W$ and $x^{\star}+r^{\star}\in W$ can be calculated as the overlap
length between the window $W$ and the shifted window $W\! -r^{\star}$,
divided by the length of $W\!$, which is $|W|=\tau$. This is correct
because the uniform distribution of points in the window 
\cite{TAO,Moody02} implies that
the frequency of any configuration is proportional to the length of
the corresponding sub-window. Clearly, the length of the overlap
between these two intervals is $0$ whenever $\lvert
r^{\star}\rvert>\tau$, and otherwise decreases linearly with 
$\lvert r^{\star}\rvert$, so we get
\begin{equation}\label{eq:auto-coeff}
\begin{split}
   \nu (r) \, & = \, \frac{\bigl| W\cap (W\! -r^{\star})\bigr|}
                      {\bigl| W\bigr|}\\[1mm] 
   & = \, \begin{cases}
   1-\frac{|r^{\star}|}{\tau} ,  & 
   \text{if $r\in\ZZ[\tau]$ and 
   $\lvert r^{\star}\rvert\leqslant \tau$,}\\
    0 , & \text{otherwise.}
   \end{cases}
\end{split}
\end{equation}
Consequently, the averaged shelling numbers for the Fibonacci 
point set are given by 
\[
  a(r)\, =\, \begin{cases} 1, &\text{if $r=0$,}\\
   2\bigl(1-\frac{|r^{\star}|}{\tau}\bigr), & \text{if
   $r\in\ZZ[\tau]$ with $|r^{\star}|\leqslant \tau$,}\\ 
   0, & \text{otherwise}.
   \end{cases} 
\]
Note that $a(r)$, for $r\in\ZZ[\tau]$, is a simple function of
$r^{\star}$, but that it behaves rather erratically if one looks at it
as a function of $r$; compare Figure~\ref{fig:fiboshell}.  The reason
behind this observation is the total discontinuity of the
\mbox{$\star$-map} from physical to internal space.

For the one-dimensional example at hand, the numbers $\nu(r)$ are
nothing but the \emph{relative probabilities} to find two points at a
distance $r$, and thus the (relatively normalised)
\emph{autocorrelation coefficients} of the point set $\vL$. As such,
they are intimately connected to the diffraction of this point
set. Clearly, correlations are much easier to handle in internal
space, where we can calculate them via volumes of intersections of
windows, as we shall see shortly.

\section{Standard approach to diffraction}\label{sec:standard}

Here, we start with a brief summary of the derivation of the
diffraction spectrum for the Fibonacci point set
$\vL=\vL_{a}\cup\vL_{b}$, considered as a cut and project set
$\vL=\{x\in L : x^{\star}\in W\}$ with $W=W_{a}\cup W_{b}$. Assume
that we place point scatterers of unit scattering strength at all
points $x\in\vL$, and consider the corresponding \emph{Dirac comb}
\[
 \omega\, =\, \delta^{}_{\! \vL} \,\defeq\, \sum_{x\in\vL} \delta^{}_{x}.
\]
We associate to $\omega$ the autocorrelation
$\gamma = \omega \circledast \widetilde{\omega}$, where
$\widetilde{\omega}= \delta^{}_{- \vL}$ is the `flipped-over'
(reflected) version of $\omega$ and $\circledast$ denotes
volume-averaged (or Eberlein) convolution \cite[Sec.~8.8]{TAO}. The
diffraction measure $\widehat{\gamma}$ is the Fourier transform of the
autocorrelation.

From the general diffraction theory for cut and project sets with
well-behaved windows, we know that the diffraction measure of this
system is a pure point measure, so consists of Bragg peaks only.
These Bragg peaks are located on the projection of the entire
\emph{dual lattice}
\[
  \cL^{*}\, =\, \myfrac{1}{\sqrt{5}} 
  \bigl\{m (\tau-1,\tau) + n (1,-1) : m,n\in\ZZ\bigr\}
\]
to the physical space (the first coordinate), which is
$L^{\circledast}=\frac{1}{\sqrt{5}} \ZZ[\tau]$. We call this set the
\emph{Fourier module} of the Fibonacci point set; it coincides with
the dynamical spectrum (in additive notation) in the mathematical
literature. Note that $\frac{1}{\sqrt{5}}=\frac{1}{5} (2\tau-1)$,
hence $L^{\circledast}\subset\QQ(\tau)$, which means that the
$\star$-map is well defined for all $k\in L^{\circledast}$. The
Fourier module is a dense subset of $\RR$, which means that the
diffraction consists of Bragg peaks on a dense set in space, where the
intensities are locally summable.

\begin{figure}
\includegraphics[width=\columnwidth]{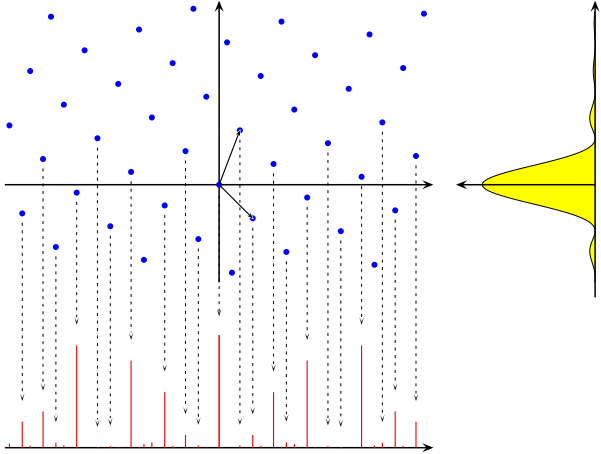}
\caption{Schematic construction of the diffraction measure of the
  Fibonacci point set from the dual lattice $\cL^{*}$ (blue dots). A
  point $(k,k^{\star})\in\cL^{*}$ results in a Bragg peak at
  $k\in L^{\circledast}$ of intensity given by the value of the
  function on the right-hand side evaluated at $k^{\star}$. Note that
  some Bragg peaks may be extinct, if the intensity function vanishes
  at $k^{\star}$.\label{fig:fibodiff}}
\end{figure}

The diffraction measure is thus the countable sum
\[
    \widehat{\gamma} \, = \sum_{k\in L^{\circledast}} \lvert
    A(k)\rvert^2 \, \delta^{}_{k}
\]
where the diffraction amplitudes, or \emph{Fourier--Bohr} (FB)
coefficients, are given by the general formula
\begin{equation}\label{eq:genampli}
   A(k) \, = \, \frac{\dens(\vL)}{\vol(W)} \, 
  \widehat{1^{}_{W}}(-k^{\star}) \, = \, 
\frac{\dens(\vL)}{\vol(W)} \, 
  \widecheck{1^{}_{W}}(k^{\star})
\end{equation}
for all $k\in L^{\circledast}$, and vanish otherwise.  Here,
\[
\begin{split}
   \widehat{g}(k) \, & = \, \int_{-\infty}^{\infty} 
   \ee^{-2\pi\ii kx}\, g(x)\dd x  \quad \text{and}\\[1mm]
   \widecheck{g}(k) \, & = \, \int_{-\infty}^{\infty} 
   \ee^{2\pi\ii kx}\, g(x)\dd x  
\end{split}
\]
denote the \emph{Fourier} and the \emph{inverse Fourier transform} of
a (here always real-valued) $L^{1}$-function $g$.  With
$\dens(\vL)=\frac{1}{5} (\tau+2)$ and $\vol(W)=\lvert W\rvert = \tau$,
Eq.~\eqref{eq:genampli} evaluates to
\[
\begin{split}
   A(k) \, &= \, \frac{1}{\sqrt{5}} 
  \int_{-1}^{\tau-1} \ee^{2\pi\ii k^{\star}y}\dd y\\[1mm]
    &=\, 
  \frac{\tau}{\sqrt{5}}\,\ee^{\pi \ii k^{\star}(\tau-2)}\, 
   \sinc(\pi \tau k^{\star})
\end{split}
\]
where $\sinc(x)=\sin(x)/x$. Hence, the diffraction intensities
are
\begin{equation}\label{eq:fibointens} 
    I(k) \, = \, \lvert A(k)\rvert^2 \, = \, 
    \biggl(\frac{\tau}{\sqrt{5}}\, \sinc(\pi \tau k^{\star})\biggr)^2
\end{equation}
for all $k\in L^{\circledast}$, and $0$ otherwise. This is illustrated
in Figure~\ref{fig:fibodiff}. Note that $I(k)$ can vanish for some
$k\in L^{\circledast}$, in which case we talk of an \emph{extinction}
of the Bragg peak. For the Fibonacci system, this may happen for
specific choices of the scattering strengths (such as in our simple
case, where we chose them to be $1$ for all points in $\vL$).
However, for a generic choice of weights (compare \eqref{eq:Fibo-int}
below), there will be no extinctions, and we will have a Bragg peak
for all $k\in L^{\circledast}$.

The corresponding autocorrelation measure $\gamma$ can be 
expressed in terms of the (dimensionless) 
\emph{pair correlation coefficients}
\[
    \nu(r) \, \defeq \, 
    \frac{\dens\bigl(\vL\cap(\vL -r)\bigr)}
    {\dens(\vL)} \, = \, \nu(-r) \ts ,
\]
which are positive for all $r\in\vL-\vL\subset \ZZ[\tau]$ and vanish
for all other distances $r$. These are precisely the coefficients we
defined in Eq.~\eqref{eq:auto-coeff} to compute the shelling
numbers. The link between the two expressions is provided by the
$\star$-map and the uniform distribution of $\vL^{\star}$ in the
window $W\!$. In terms of these pair correlation coefficients, the
autocorrelation measure is
\[
    \gamma \, = \, \dens(\vL) \!
    \sum_{r\in\vL-\vL} \! \nu(r)\, \delta_{r},
\]
which is a pure point measure supported on the difference set
$\vL-\vL$.

More generally, we may associate two different, in general complex,
scattering strengths $u^{}_{a}$ and $u^{}_{b}$ to the points in
$\vL_{a}$ and $\vL_{b}$, respectively, and consider the weighted Dirac
comb $\omega =
u^{}_{a}\delta^{}_{\!\vL_{a}}+u^{}_{b}\delta^{}_{\!\vL_{b}}$. In this
case, the diffraction intensity for all wave numbers
$k\in L^{\circledast}$ is given by the superposition
\begin{equation}\label{eq:Fibo-int}
  I(k) \, = \, 
    \bigl|  u^{}_{a} \, A^{}_{a}(k) + 
               u^{}_{b} \, A^{}_{b}(k)\bigr|^{2}
\end{equation}
of the corresponding FB amplitudes 
\begin{align*}
  A^{}_{a,b}(k) \, &= \, \frac{\dens(\vL_{a,b})}{\vol(W_{\! a,b})} \,
  \widehat{1^{}_{W_{\! a,b}}}(-k^{\star})\\[1mm]
  & = \,  \frac{\dens(\vL)}{\vol(W)} \,
  \widehat{1^{}_{W_{\! a,b}}}(-k^{\star}) \, = \, 
  \frac{1}{\sqrt{5}}  \, \widehat{1^{}_{W_{\! a,b}}}(-k^{\star}).
\end{align*}
The corresponding autocorrelation measure can once more be expressed
in terms of pair correlation functions, now distinguishing points in
$\vL_{a}$ and $\vL_{b}$,
\[
    \nu^{}_{\alpha\beta}(r) \, \defeq \, 
    \frac{\dens\bigl(\vL_{\alpha}\cap(\vL_{\beta} -r)\bigr)}
    {\dens(\vL)} \, = \, \nu^{}_{\beta\alpha}(-r) \ts .
\]
These coefficients are positive for all $r\in\vL_{\beta}-\vL_{\alpha}$
and vanish otherwise, and in particular satisfy the relation
$\sum_{\alpha,\beta\in\{a,b\}}\nu^{}_{\alpha\beta} (r) = \nu(r)$.

The relation \eqref{eq:genampli} between the FB coefficients and the
Fourier transform of the compact windows holds for any regular model
set, which is a cut and project set with some `niceness' constraint on
the window; see \cite[Thm.~9.4]{TAO} for details.  While this works
well for many of the nice examples with polygonal windows, it becomes
practically impossible to compute the FB coefficients in this way if
the windows are compact sets with fractal boundaries. Such windows
naturally arise for cut and project sets which also possess an
inflation symmetry. Indeed, some of the structure models of
icosahedral quasicrystals, see \cite{CdYb} for an example, feature
experimentally determined windows whose shapes may indicate first
steps of a fractal construction of the boundary.

Let us therefore explain a different approach that will permit an
efficient computation of the diffraction also for such, more
complicated, situations.

\section{Renormalisation and internal cocycle}\label{sec:cocycle}

Let us reconsider our motivating example, the Fibonacci point sets
$\vL^{}_{a,b}$ of Eq.~\eqref{eq:vL-ab}. We will use both their
inflation structure and their description as cut and project
sets. Here, we make use of the iteration \eqref{eq:vL-rec} and the
corresponding relation \eqref{eq:W-rec} for the windows (or, more
precisely, the closure of the windows).  This inflation structure
induces the following relation between the characteristic functions of
the windows,
\begin{equation}\label{eq:charfun-rec}
    1^{}_{W_{\! a}} \, =\,  1^{}_{\sigma W_{\! a}\cup\ts\sigma W_{b}} 
    \quad\text{and}\quad
    1^{}_{W_{b}} \, = \, 1^{}_{\sigma W_{\! a} +\ts\sigma} \ts ,
\end{equation}
where we again set $\sigma=\tau^{\star}=1-\tau$.  Since the (closed)
windows only share at most boundary points, we observe that
$1^{}_{\sigma W_{\! a}\cup\ts\sigma W_{b}} =1^{}_{\sigma W_{\! a}} +
1^{}_{\sigma W_{b}}$ holds as an equality of $L^{1}$-functions. Now,
we can apply the Fourier transform, where it will turn out to be more
convenient to work with the \emph{inverse} Fourier transform from the
start. Applying this transform yields the relations
\begin{equation}\label{eq:charfun}
   \widecheck{1^{}_{W_{\! a}}} \, = \:
   \widecheck{1^{}_{\sigma W_{\! a}}} + \,
     \widecheck{1^{}_{\sigma W_{b}}}
    \quad\text{and}\quad
   \widecheck{1^{}_{W_{b}}} \, = \:
   \widecheck{1^{}_{\sigma W_{\! a}+\ts \sigma}}\ts .
\end{equation}
These equations capture the action of the inflation in internal space
in terms of functional equations for the inverse Fourier transform of
the windows, which in turn determine the diffraction.  Note that, by
an elementary change of variable calculation in the Fourier integral,
one has
\begin{equation}\label{eq:affineFT}
    \widecheck{1^{}_{\alpha K+\beta}}(y) \, = \, 
    \lvert\alpha\rvert \, \ee^{2\pi\ii \beta y} \, 
    \widecheck{1^{}_{K}}(\alpha\ts y)
\end{equation}
for arbitrary $\alpha,\beta\in\RR$ with $\alpha\ne 0$ and any compact
set $K\subset \RR$. This can be used to express the functions in
\eqref{eq:charfun} with $\sigma$-scaled and shifted windows in terms
of the indicator functions of the original windows.

Indeed, defining 
\begin{equation}\label{eq:def-h}
  h^{}_{a,b}\defeq\widecheck{1^{}_{W_{\! a,b}}}
\end{equation}
for the two functions involving the original windows, and using
Eq.~\eqref{eq:affineFT}, we can rewrite Eq.~\eqref{eq:charfun} as
\begin{equation}\label{eq:self}
   \begin{pmatrix} h^{}_{a} \\ h^{}_{b} \end{pmatrix} (y) \, = \, 
   \lvert\sigma\rvert  \, \sB (y) 
   \begin{pmatrix} h^{}_{a} \\ h^{}_{b} \end{pmatrix} (\sigma y)
\end{equation}
with the matrix
\begin{equation}\label{eq:sB}
\sB (y) \, \defeq \, \begin{pmatrix} 1 & 1 \\ 
    \ee^{2\pi\ii\sigma y} & 0 \end{pmatrix}.
\end{equation}
The matrix $\sB$ is obtained by first taking the $\star$-map of the
set-valued displacement matrix $T$ of Eq.~\eqref{eq:T} and then its
inverse Fourier transform. For this reason, $\sB$ is called the
\emph{internal Fourier matrix} \cite{BG19c}, to distinguish it from the
Fourier matrix of the renormalisation approach in physical space
\cite{BG16,BFGR}; see \cite{BS18,BS20} for various extensions
with more flexibility in the choice of the interval lengths.

In Dirac notation, we set $|\ts h\rangle=(h^{}_{a},h^{}_{b})^T$, which
satisfies $|\ts h(0)\rangle=\tau \ts |\ts v\rangle$ with the right
eigenvector $ |\ts v\rangle$ of the substitution matrix $M$ from
Eq.~\eqref{eq:ev}. Applying the iteration \eqref{eq:self} $n$ times
then gives
\begin{equation}\label{eq:h-rec}
    |\ts h(y)\rangle \, = \, \lvert\sigma\rvert^{n} \sB^{(n)}(y) \,
    |\ts h(\sigma^{n} y)\rangle 
\end{equation}
where
\[
    \sB^{(n)}(y)\, \defeq \, \sB (y) \sB (\sigma y) \cdots 
    \sB (\sigma^{n-1} y)\ts .
\]
In particular, these matrices satisfy $\sB^{(1)}=\sB$ and
$\sB^{(n)}(0)=M^{n}$ for all $n\in\NN$, where $M$ is the substitution
matrix from Eq.~\eqref{eq:fibmat}, as well as the relations
\begin{equation}\label{eq:Bsplit}
   \sB^{(n+m)}(y) \, = \, \sB^{(n)}(y) \,\sB^{(m)}(\sigma^n y)
\end{equation}
for any $m,n\in\NN$. Note that $\sB^{(n)}(y)$ defines a matrix
cocycle, called the \emph{internal cocycle}, which is related to the
usual inflation cocycle (in physical space) by an application of the
$\star$-map to the displacement matrices of the powers of the
inflation rule; compare \cite{BGM19,BG19c} and see \cite{BS18,BS20}
for a similar approach. Note also that $\lvert\sigma\rvert <1$, which
means that $\sigma^n$ approaches $0$ exponentially fast as
$n\to\infty$. We can exploit this exponential convergence to
efficiently compute the diffraction amplitudes, which are proportional
to the elements of the vector $|\ts h\rangle$.

Considering the limit as $n\to\infty$ in Eq.~\eqref{eq:h-rec}, one can
show that 
\begin{equation}\label{eq:h}
   |\ts h(y)\rangle=C(y) |\ts h(0)\rangle
\end{equation}
with
\begin{equation}\label{eq:C}
   C(y) \, \defeq  \lim_{n\to\infty} 
   \lvert\sigma\rvert^n \sB^{(n)}(y)\ts ,
\end{equation}
which exists pointwise for every $y\in\RR$. In fact, one has compact
convergence, which implies that $C(y)$ is continuous \cite[Thm.~4.6
  and Cor.~4.7]{BG19c}.  Clearly, since $\sB^{(n)}(0)=M^n$, we have
$C(0)=P$ with the projector $P=|\ts v\rangle\langle u\ts |$ from
Eq.~\eqref{eq:Pdef}.

Using Eq.~\eqref{eq:Bsplit} with $m=1$
and letting $n\to\infty$, one obtains
\[
   \tau \, C(y) \, = \, C(y) M\ts ,
\]
since $\lvert\sigma\rvert=\tau^{-1}$.  This relation implies that each
row of $C(y)$ is a multiple of the left eigenvector $\langle u|$ of
the substitution matrix $M$ from Eq.~\eqref{eq:ev}, so there is
a vector-valued function $|\ts c(y) \rangle$ such that
\begin{equation}\label{eq:Cu}
  C(y)\, =\, |\ts c(y)\rangle\langle u\ts |
\end{equation}
holds with
$|\ts c(y)\rangle = \bigl(c^{}_{a}(y),c^{}_{b}(y)\bigr)^T \nts$, where
we have $|\ts c(0)\rangle=|\ts v\rangle$.

From Eqs.~\eqref{eq:h} and \eqref{eq:Cu}, we obtain 
\[ 
 |\ts h(y)\rangle\, =\, |\ts c(y)\rangle\langle u\ts 
 |\ts h(0)\rangle\, =\, \tau \ts\ts |\ts c(y)\rangle
\]
and the inverse Fourier transforms of the windows from 
Eq.~\eqref{eq:def-h} are thus encoded in the matrix $C$. 

For the Fibonacci case, we can calculate $|\ts c(y)\rangle$ by taking
the Fourier transforms of the known windows $W^{}_{\! a,b}$ to obtain
\begin{align*}
     c^{}_{a}(y) \, &= \, 
     \myfrac{\ee^{2\pi\ii (\tau-1)y} - \ee^{2\pi\ii (\tau-2)y}}
     {2\pi\ii y}
\intertext{and}
     c^{}_{b}(y) \, &= \, 
     \myfrac{\ee^{2\pi\ii (\tau-2)y} - \ee^{-2\pi\ii y}}
     {2\pi\ii y} \ts .
\end{align*}
Note that these functions never vanish simultaneously, so $C(y)$ is
always a matrix of rank $1$. However, taking the Fourier transform of
the windows takes us essentially back to the standard approach. 

The main benefit of the internal cocycle approach is that it applies
in other situations, where no explicit calculation of the (inverse)
Fourier transform of the windows is feasible. This is achieved via
\emph{approximating} $C(y)$ by $\lvert \sigma\rvert^n \sB^{(n)}(y)$
for a sufficiently large $n$, such that $\lvert \sigma\rvert^n y$ is
small and $C(y)$ is approximated sufficiently well. This works because
the (closed) windows are compact sets, so that their (inverse) Fourier
transforms are continuous functions. The convergence of this
approximation is exponentially fast. We refer to \cite{BG19c} for
further details and an extension of the cocycle approach to more
general inflation systems, and to \cite{BG20} for a planar example.

From the general formula \eqref{eq:genampli} for regular model sets,
the FB amplitudes are
\begin{equation}\label{eq:FB-h}
   A^{}_{\!\vL_{ a,b}}(k) \, = \, 
   \frac{h^{}_{a,b}(k^{\star})}{\sqrt{5}} \, = \, 
   \frac{\tau}{\sqrt{5}}\, c^{}_{a,b}(k^{\star})
\end{equation}
for $k\in L^{\circledast}$.  So, the relevant input is the knowledge
of the Fourier module, which determines where the Bragg peaks are
located. Then, one can approximate $C$ by evaluating the matrix
product in Eq.~\eqref{eq:C}, for any chosen $k\in L^{\circledast}$,
at $y = k^{\star}$ and with a
sufficiently large $n$. In what follows, numerical calculations and
illustrations are based on this cocycle approach due to its superior
speed and accuracy in the presence of complex windows.

\section{Fractally bounded windows}\label{sec:planar}

The internal cocycle approach of Section~\ref{sec:cocycle} was first
applied to a ternary inflation tiling with the smallest
Pisot--Vijayaraghavan (PV) number (also known as the `plastic number')
as its inflation multiplier \cite{BG20}. In the cut and project
description, the internal space of this one-dimensional tiling is
two-dimensional, and the windows are \emph{Rauzy fractals}
\cite{PF}. This means that the windows are still topologically
regular, so each window is the closure of its interior, but have a
fractal boundary of zero Lebesgue measure.  Consequently, the general
diffraction result for model sets still applies, and the diffraction
is given by the Fourier transform of the windows as described
above. In turn, this means that the internal cocycle approach applies
and can be used to compute the Fourier transforms and the diffraction
intensities for such tilings; see \cite{BG20} for details.

\begin{figure}[t]
\includegraphics[width=0.7\columnwidth]{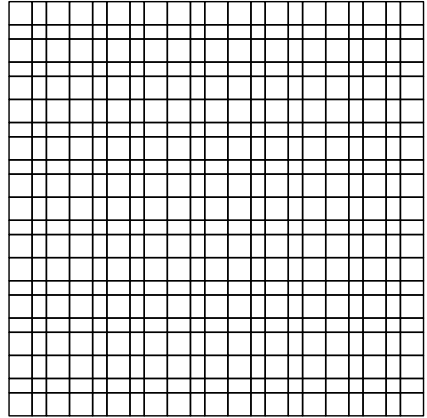}
\caption{Patch of the square Fibonacci tiling.\label{fig:fibosq}}
\end{figure}

Here, we discuss examples of planar projection tilings with fractally
bounded windows, which are based on \emph{direct product variations}
(DPVs) \cite{Sadun,Frank15} of Fibonacci systems, as recently
described in \cite{BFG21}. Clearly, if one considers a direct product
structure based on the Fibonacci tiling, one obtains a tiling of the
plane, called the \emph{square Fibonacci tiling}. This tiling has been
used as a toy model for the study of electronic properties
\cite{Ron,Mandel,David}, but has been observed experimentally to form
in a molecular overlayer on a two-fold surface of an icosahedreal
quasicrystal \cite{CSMS}.  It is built from four prototiles, a large
square of edge length $\tau$, a small square of edge length $1$, and
two rectangles with a long ($\tau$) and a short ($1$) edge; see
Figure~\ref{fig:fibosq}.

\begin{figure}[b]
\includegraphics[width=0.7\columnwidth]{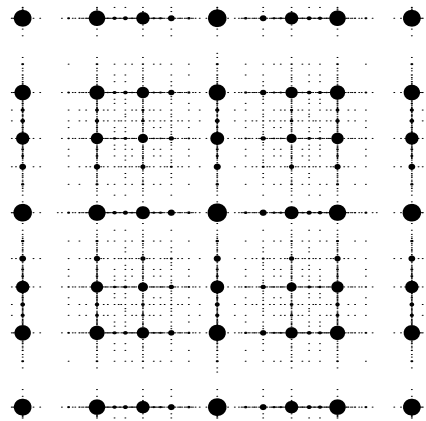}
\caption{Central part of the diffraction image of the square Fibonacci
  tiling.\label{fig:fibosqdiff}}
\end{figure}

As a direct product of inflation tilings, this two-dimensional
square Fibonacci tiling also possesses an inflation rule, which takes
the form
\begin{equation}\label{eq:fibosqrule}
  \raisebox{-0.1\columnwidth}{\includegraphics[width=0.6\columnwidth]{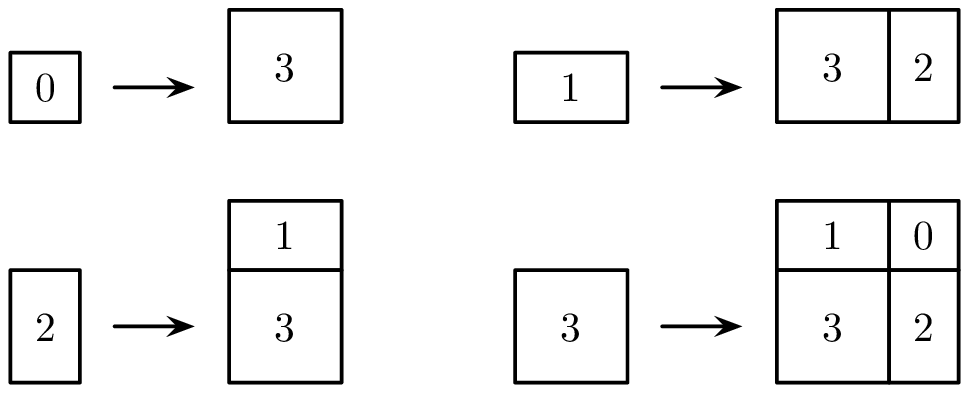}}
\end{equation}
where we labelled the small and large squares by $0$ and $3$, and the
two rectangles by $1$ and $2$, respectively. A DPV is now obtained by
modifying these rules while keeping the stone inflation character
intact, thus probing the ideas of \cite{CS} into a slightly different
direction. Clearly, there are two possibilities to rearrange the
images of the rectangles by swapping the two tiles, and a close
inspection shows that there are altogether $12$ ways of rearranging
the image of the large square. This means that there are $48$ distinct
inflation rules in total, which all share these prototiles and the
same inflation matrix.

Due to the direct product structure, the square Fibonacci tiling
clearly possesses a cut and project description. The windows for the
four prototiles are obtained as products of the original windows. The
product structure thus extends to the diffraction measure, which is
supported on the Fourier module 
\[
  L^{\circledast}\nts\times\ts L^{\circledast}\ts ,
\]
where $L^{\circledast}\, =\, \frac{1}{\sqrt{5}}\ZZ[\tau]$ is the
Fourier module of the one-dimensional Fibonacci tiling. The
diffraction amplitudes are also given by products of those for the
one-dimensional system, and are thus easy to compute. An illustration
of the diffraction pattern is shown in Figure~\ref{fig:fibosqdiff}.
Here, Bragg peaks are represented by disks, centred at the position of
the peak, with an area proportional to their intensities.

It turns out that \emph{all} $48$ DPV inflation tilings are regular
model sets, and hence are pure point diffractive; see
\cite[Thm.~5.2]{BFG21}. They all share the same Fourier module,
$L^{\circledast}\times L^{\circledast}$. This implies that the Bragg
peaks are always located at the same positions (where we disregard
possible extinctions). However, their intensities are determined by
the Fourier transform of the windows, and it turns out that the
windows of these DPVs can differ substantially.

In particular, $20$ of these DPVs possess windows of Rauzy fractal
type, of which there are three different types, called `castle',
`cross' and `island' in \cite{BFG21}. They have different fractal
dimension of the window boundaries, which are approximately $1.875$,
$1.756$ and $1.561$, respectively. As the dimensions are all smaller
than two, is it obvious that these boundaries have zero Lebesgue
measure.

\begin{figure}[t]
\includegraphics[width=0.7\columnwidth]{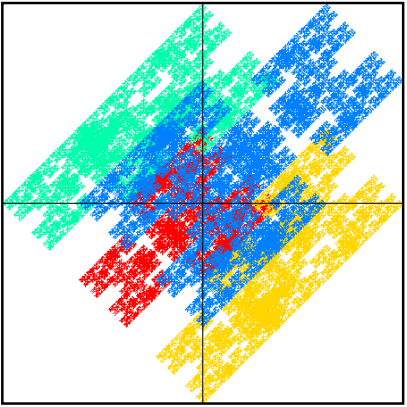}
\caption{Castle-type window for the DPV \eqref{eq:castle}. The windows
  for the four types of tiles are distinguished by colour, namely red
  ($0$), yellow ($1$), green ($2$) and blue ($3$).  The outer box
  marks the square $[-\tau, \tau]^2$, with the coordinate axes
  indicated as well.\label{fig:castle}}
\end{figure}

\begin{figure}[b]
\includegraphics[width=0.7\columnwidth]{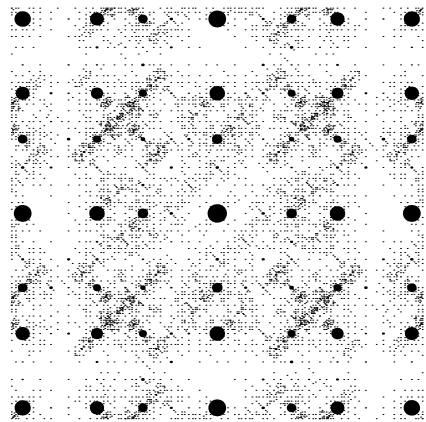}
\caption{Diffraction image of the DPV
  \eqref{eq:castle}.\label{fig:castlediff}}
\end{figure}

In what follows, we are going to illustrate some properties of these
DPVs with three examples, one for each of these fractally bounded
window types. The inflation rules for the three examples have the same
images for the small square (tile $0$) and both rectangles (tiles $1$
and $2$) as the square Fibonacci rule of Eq.~\eqref{eq:fibosqrule},
and thus only differ in the image of the large square (tile $3$). For
a discussion of the complete set of $48$ DPVs, we refer to
\cite{BFG21}.

\begin{figure}[t]
\includegraphics[width=0.7\columnwidth]{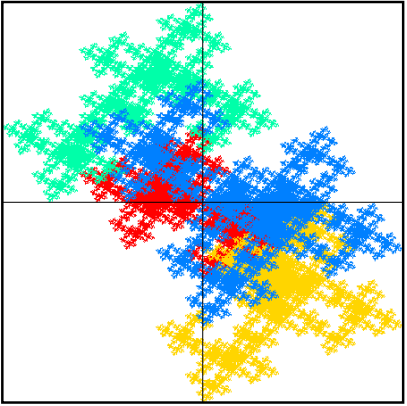}
\caption{Cross-type window for the DPV
  \eqref{eq:cross}.\label{fig:cross}}
\end{figure}

\begin{figure}[b]
\includegraphics[width=0.7\columnwidth]{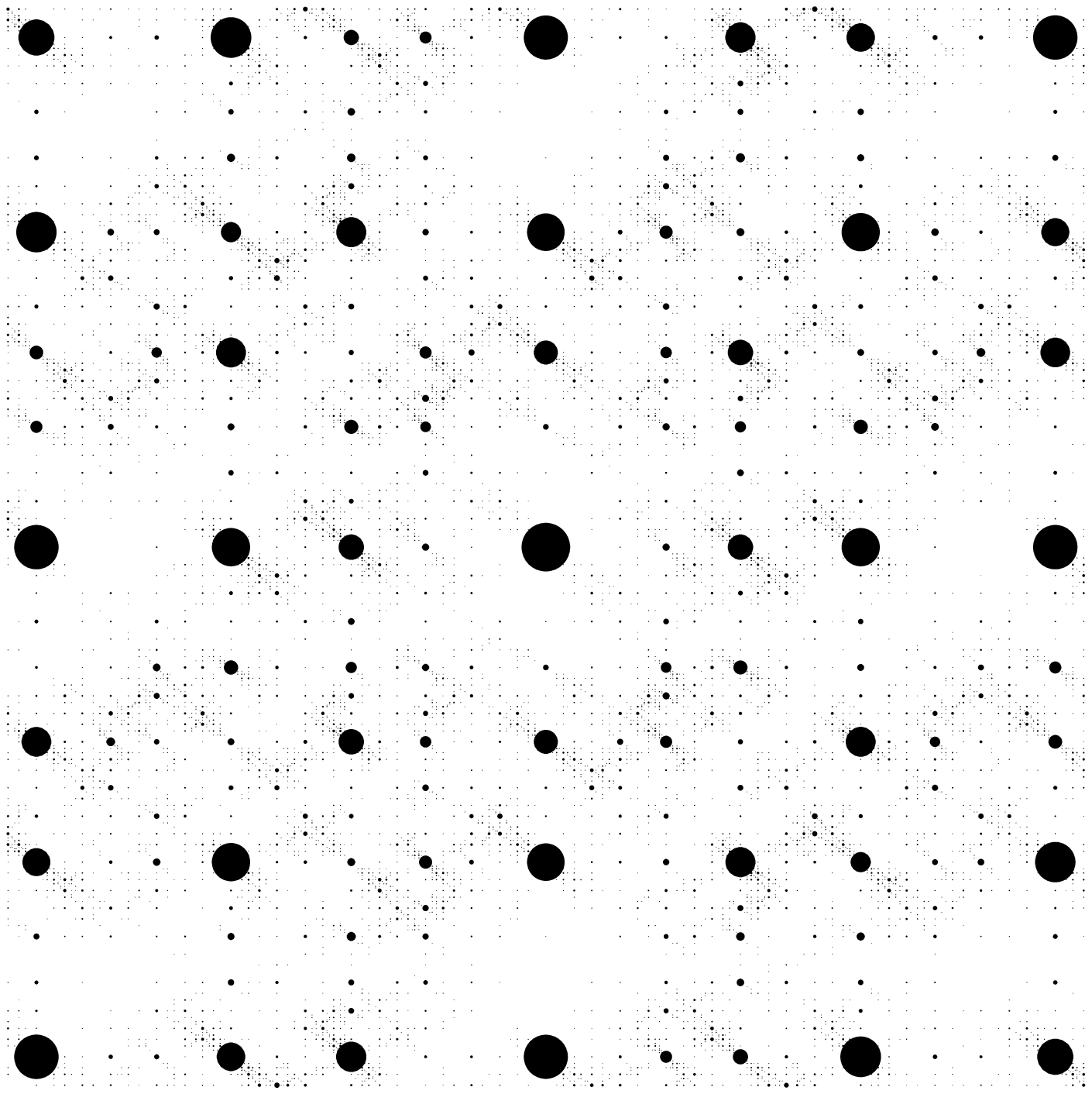}
\caption{Diffraction image of the DPV
  \eqref{eq:cross}.\label{fig:crossdiff}}
\end{figure}

For the castle-type windows of Figure~\ref{fig:castle}, we use the
inflation
\begin{equation}\label{eq:castle}
\raisebox{-0.07\columnwidth}{\includegraphics[width=0.4\columnwidth]{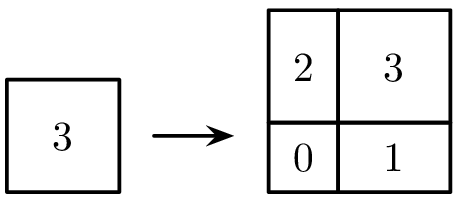}}
\end{equation}
for the large square. Note that this rule dissects the inflated large
square such that there is a reflection symmetry along the main
diagonal, which will be reflected in a symmetry of the tiling (which
maps the squares onto themselves and interchanges the
rectangles). This is also apparent for the windows in
Figure~\ref{fig:castle}. The windows for the large and small squares
are mapped onto themselves under reflection at the main diagonal,
while the windows for the rectangular tiles are interchanged. The
diffraction pattern also respects this symmetry; see
Figure~\ref{fig:castlediff}.

\begin{figure}[t]
\includegraphics[width=0.7\columnwidth]{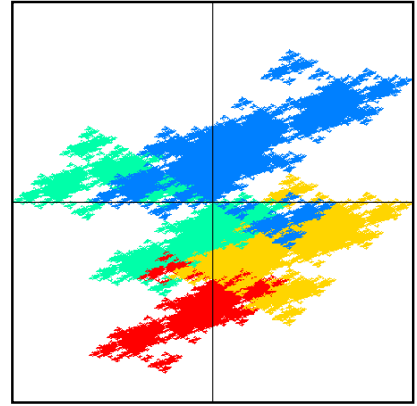}
\caption{Island-type window for the DPV \eqref{eq:island}.\label{fig:island}}
\end{figure}

\begin{figure}[b]
\includegraphics[width=0.7\columnwidth]{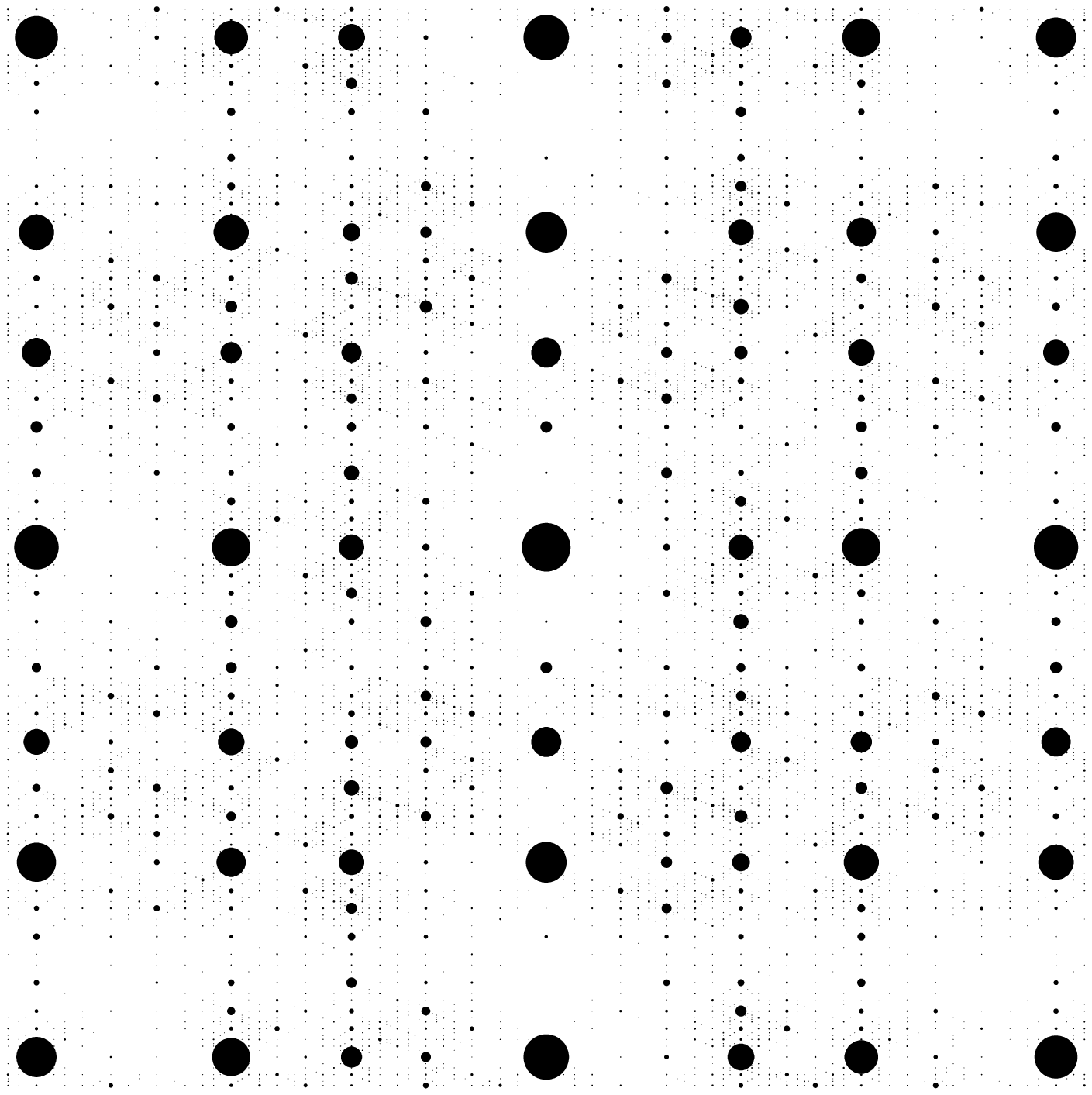}
\caption{Diffraction image of the DPV
  \eqref{eq:island}.\label{fig:islanddiff}}
\end{figure}

For the cross-type windows, the inflation of the large square is given
by
\begin{equation}\label{eq:cross}
\raisebox{-0.07\columnwidth}{\includegraphics[width=0.4\columnwidth]{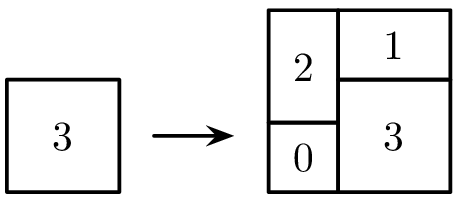}}
\end{equation}
which, in contrast to the previous example, has no reflection
symmetry. Consequently, neither the windows shown in
Figure~\ref{fig:cross} nor the diffraction image illustrated in
Figure~\ref{fig:crossdiff} have any reflection symmetry.

The same is true for the final example with the island-type window
shown in Figure~\ref{fig:island}. This corresponds to the inflation
\begin{equation}\label{eq:island}
\raisebox{-0.07\columnwidth}{\includegraphics[width=0.4\columnwidth]{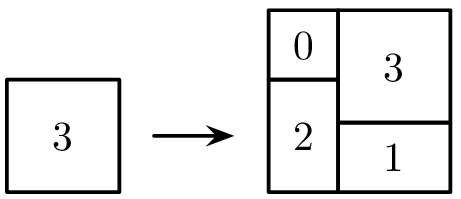}}
\end{equation}
of the large square tile. The corresponding diffraction pattern is
illustrated in Figure~\ref{fig:islanddiff}.

Comparing the diffraction patterns of Figures~\ref{fig:castlediff},
\ref{fig:crossdiff} and \ref{fig:islanddiff} with those of the square
Fibonacci tiling shown in Figure~\ref{fig:fibosqdiff}, we note that
the strongest peaks are almost unchainged, while the intensities of
the weaker peaks show some intriguing behaviour. The reason for this
behaviour is that all three model sets are subsets of a common Meyer
set, and the so-called $\varepsilon$-dual characters of the difference
set of this Meyer set, for small $\varepsilon$, always give rise to
high intensity Bragg peaks; see \cite{Str13} for details. This is the
reason why the strongest peaks stay almost the same.

For the fractally-bounded windows, one generally sees more peaks,
which is due to the larger spread of the window in internal space, and
the slower asymptotic decay of the Fourier transform of the window (as
$k^{\star}\to\infty$). With limited resolution, some of the intensity
distributions on these peaks could resemble continuous components, so
might potentially be mistaken as such in experiments.

\section{Diffraction and hyperuniformity}\label{sec:hyper}

The discovery of quasicrystals highlighted the lack of a clear
definition of the concept of \emph{order}. In crystallography,
diffraction is the main tool to detect long-range order, and a pure
point diffraction is generally associated to an ordered,
(quasi)crystalline structure, while absolutely continuous diffraction
is typically seen as an indication of random disorder (but see
\cite{Frank03,BG09,CG,CGS} for examples of deterministic structures
that show absolutely continuous diffraction). Here, we briefly discuss
a related concept that has recently gained popularity.

From the original idea of using the degree of `(hyper)uniformity' in
density fluctuations in many-particle systems \cite{TS,BGK,BGKZ} to
characterise their order, the \emph{scaling} behaviour of the total
diffraction intensity near the origin has emerged as a possible
measure to capture long-distance correlations. As far as aperiodic
structures are concerned, there are in fact a number of early, partly
heuristic, results in the literature \cite{Luck,Aubry,GL}. These have
recently been reformulated and extended \cite{Josh1,Josh2} and
rigorously established \cite{BG19b}, using exact renormalisation
relations for primitive inflation rules
\cite{BFGR,BG16,Neil,BGM19,BGM18,NeilDiss}; see also \cite{FMV} for
results for some planar aperiodic tilings.

For the investigation of scaling properties, we follow the
existing literature and define
\begin{equation}\label{eq:Z-def}
     Z (k) \, \defeq \, \widehat{\gamma} \bigl( (0,k]\bigr) ,
\end{equation}
which is a modified version of the \emph{distribution function} of the
diffraction measure. Here, $Z(k)$ is the total diffraction intensity
in the half-open interval $(0,k]$, and thus ignores the central peak.
Due to the point reflection symmetry of $\widehat{\gamma}$ with 
respect to the origin, this quantity can also be expressed as
\[
    Z (k) \, = \, \myfrac{1}{2}
    \Bigl(\widehat{\gamma} \bigl( [-k,k]\bigr) -
    \widehat{\gamma} \bigl( \{0\}\bigr) \Bigr).
\]
The interest in the scaling of $Z(k)$ as $k\to 0$ is motivated by the
intuition that the small-$k$ behaviour of the diffraction measure
probes the long-wavelength fluctuations in the structure. As the
latter is related to the variance in the distribution of patches, it
can serve as an indicator for the degree of uniformity of the
structure \cite{TS}.  It is obvious that any periodic structure leads
to $Z(k) = 0$ for all sufficiently small wave numbers $k$.

Here, we review the result for variants of the one-dimensional
Fibonacci model sets considered above, where we now allow for changes
of the windows. For a general discussion of this approach and more
examples of systems with different types of diffraction, we refer to
\cite{BG19b} and references therein.

Let us look at the diffraction for a cut and project set with the same
setup as the Fibonacci tiling considered in
Section~\ref{sec:standard}, but with the window $W$ replaced by an
arbitrary finite interval of length $s$. Note that these tilings, in
general, do \emph{not} possess an inflation symmetry. Nevertheless,
the diffraction intensity is still of the form \eqref{eq:fibointens},
but now featuring the interval length $s$, and is given by
\[
    I(k) \, = \, I(0) \bigl(\sinc(\pi s k^{\star})\bigr)^2
\]
for all $k\in L^{\circledast}$.  Now, consider a sequence of positions
$\tau^{-\ell} k$ with $k \in L^{\circledast}$ and
$\ell\in\Nnull$. Since we have $\sinc (x) = \sin(x)/x = \cO (x^{-1} )$
as $x \to\infty$, it follows that $I ( \tau^{-\ell}k ) = \cO \bigl(
\tau^{-2\ell} \bigr)$ as $\ell\to\infty$.

Consequently, the sum of intensities along the series of peaks,
\[
  \vS (k)\, =\, \sum_{\ell=0}^{\infty} I(\tau^{-\ell}k)\ts ,
\]
satisfies the asymptotic behaviour
\[
   \vS (\tau^{-\ell} k) \, \sim\, c(k)\, \tau^{-2\ell}\, \vS (k)
\]
as $\ell\to\infty$, where it can be shown that $c(k) = \cO(1)$
\cite{BG19b}. Expressing $Z(k)$ in terms of these sums gives
\[
   Z(k) \; = \! \sum_{\substack{\;\kappa\in L^{\circledast} \\
       \frac{k}{\tau} < \kappa \leqslant k }} \!\! \vS (\kappa )\ts ,
\]
which, for $\ell\to\infty$,  implies the asymptotic behaviour
\[
   Z(\tau^{-\ell} k) \, \asymp \, \tau^{-2\ell}\, Z(k)\ts , 
\]
where the implied constants may still depend on $k$, but are $\cO (1)$
as $k\,\raisebox{2pt}{$\scriptscriptstyle \searrow$}\, 0$.  This leads
to a power-law scaling behaviour of the form $Z(k) = \cO (k^2)$ as
$k\,\raisebox{2pt}{$\scriptscriptstyle \searrow$}\, 0$.

This generic result remains true if we choose a window which
corresponds to a tiling with inflation symmetry, which requires the
window to be an interval of length $s\in\ZZ[\tau]$. This obviously
holds for our original Fibonacci window $W$ of length $\tau$.
However, one gets a stronger result for this case \cite{BG19b,Josh1},
as we shall now recall.

Choosing $s\in\ZZ[\tau]$ means $s=a+b \tau$ with $a,b\in\ZZ$. For
$0\ne k\in L^{\circledast}$, set $k = \kappa/\mbox{\small $\sqrt{5}$}$
with $\kappa = m + n \tau$ for some $m,n \in \ZZ$, excluding
$m=n=0$. Applying the $\star$-map then gives
\[
    I(\tau^{-\ell} k ) \, = \, I(0) \,
    \biggl(\sinc \Bigl( \myfrac{\pi \tau^{\ell} s \ts 
        \kappa^{\star}}{\mbox{\small $\sqrt{5}$}}
    \Bigr)\biggr)^2 ,
\] 
with $\ell\in\NN_0$. 

Now, denote by $f_n$ with $n\in\ZZ$ the Fibonacci numbers defined by
$f^{}_{0} = 0$, $f^{}_{1} = 1$ and the recursion
$f^{}_{n+1} = f^{}_{n} + f^{}_{n-1}$. They satisfy the well-known
formula
\begin{equation}\label{eq:fib-form}
      f^{}_n \, = \, \myfrac{1}{\mbox{\small $\sqrt{5}$}} \,
      \Bigl( \tau^n - \bigl( -1/\tau\bigr)^n \Bigr) 
\end{equation}
for all $n\in\ZZ$. Using this relation, 
we obtain
\begin{equation}\label{eq:fiboexp}
\begin{split}
   \sin \Bigl( \myfrac{\pi \tau^{\ell} s \ts \kappa^{\star}}
    {\mbox{\small $\sqrt{5}$}}
    \Bigr)^2   &  = \:
    \sin \Bigl( \myfrac{\pi \ts \lvert s \ts 
         \kappa^{\star} \rvert}{\mbox{\small
    $ \sqrt{5}$}} \, \tau^{-\ell}  \Bigr)^2  \\[1mm]
    & = \, \myfrac{\pi^2 ( s \ts \kappa^{\star} )^2}{5}
      \, \tau^{-2\ell}  \, + \cO \bigl( \tau^{-6 \ell}\bigr) 
\end{split}
\end{equation}
as $\ell\to\infty$. Here, 
the first step follows by using Eq.~\eqref{eq:fib-form}
to replace $\tau^{\ell}/\mbox{\small $\sqrt{5}$}$
and then reducing the argument via the relation
\[
     \sin (m \pi + x) \, = \, (-1)^m \sin (x)\ts ,
\]
which holds for all $m\in\ZZ$ and $x\in\RR$.  This is possible because
all Fibonacci numbers are integers. The second step then uses the
Taylor approximation $\sin (x) = x + \cO (x^3)$ for small values of
$x$.

\begin{figure}
\centerline{\includegraphics[width=0.9\columnwidth]{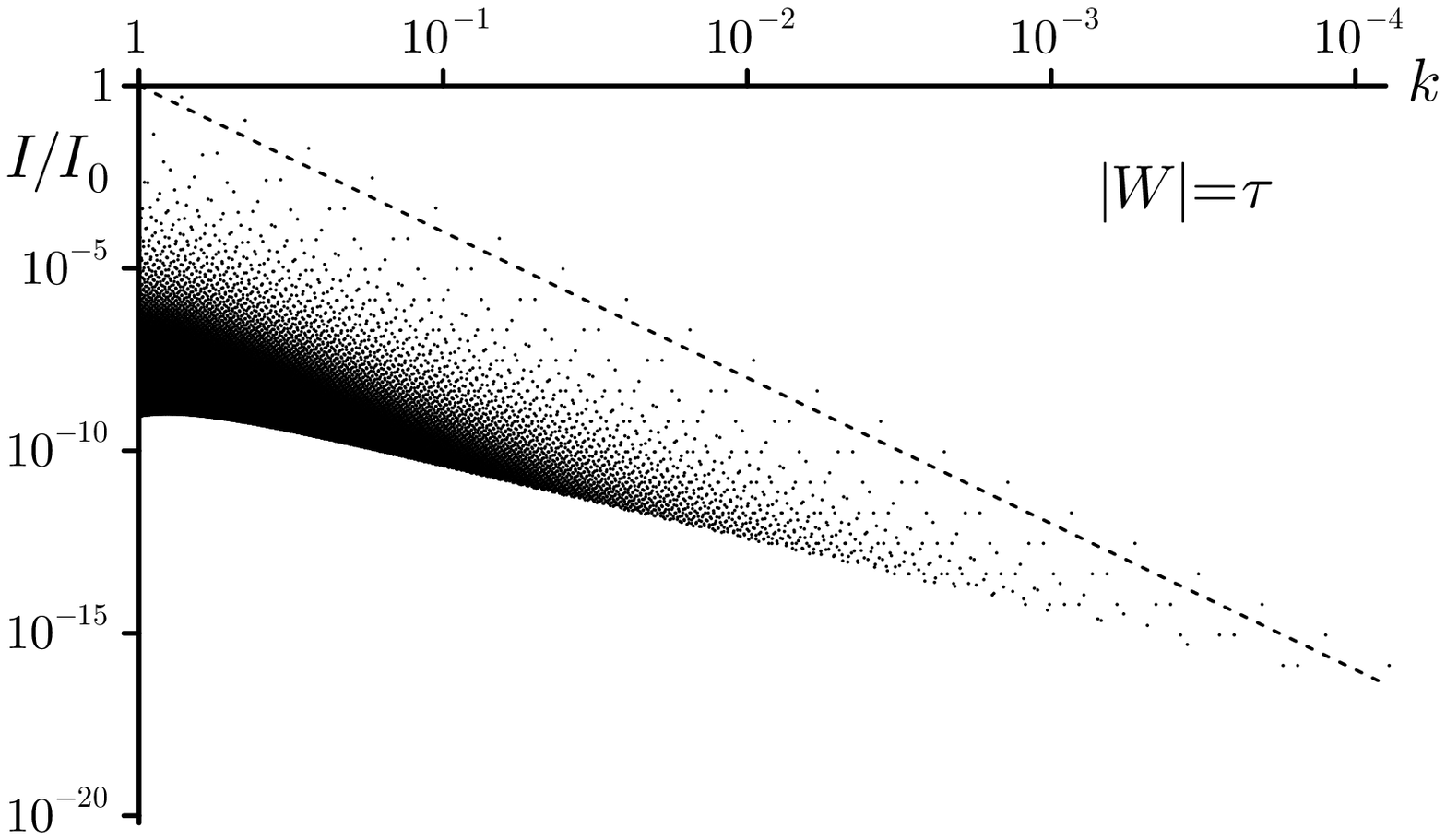}}
\centerline{\includegraphics[width=0.9\columnwidth]{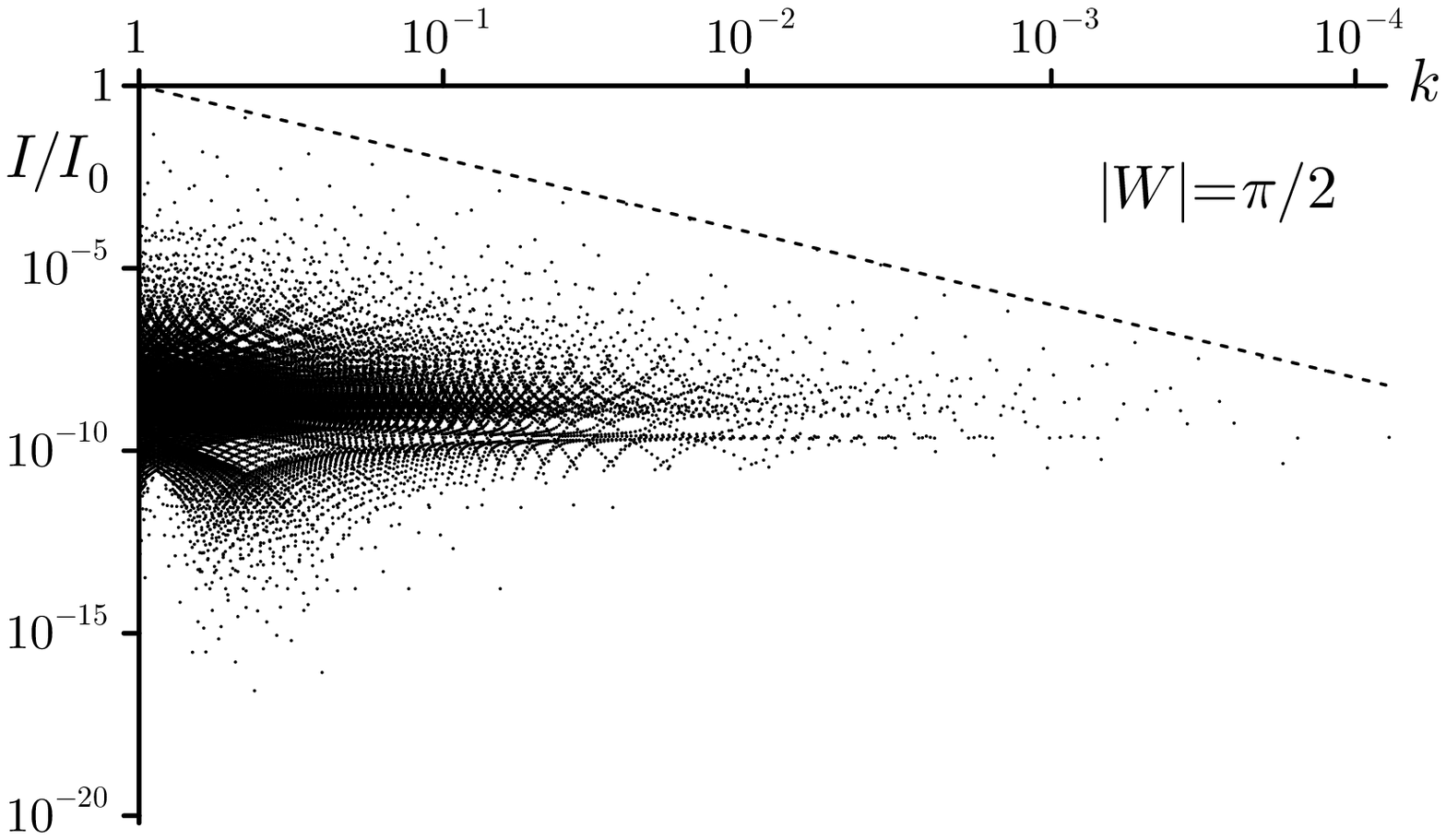}}
\centerline{\includegraphics[width=0.9\columnwidth]{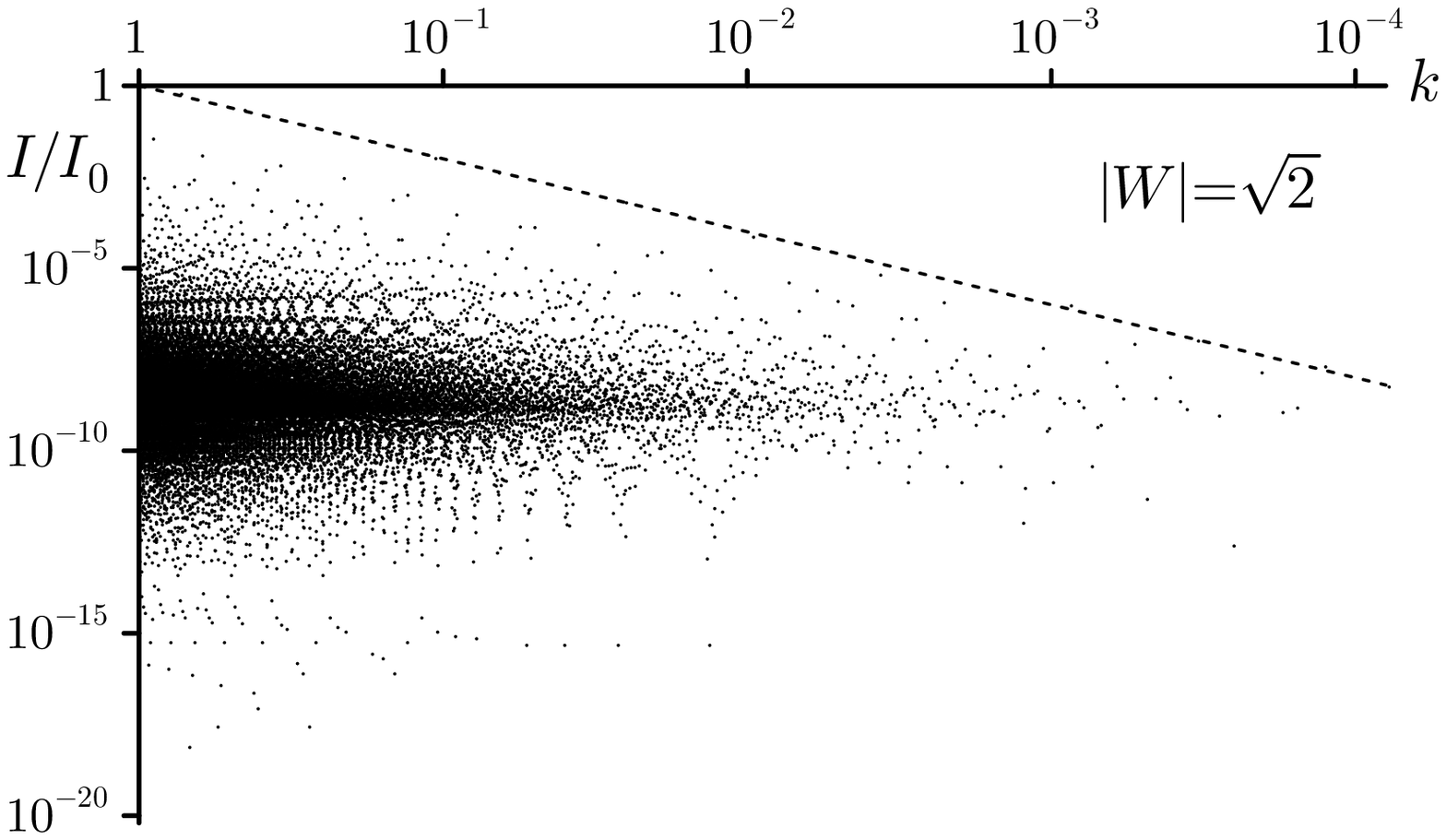}}
\caption{Double logarithmic plot of the intensity ratio $I/I_{0}$ of
  Bragg peaks located at $k=(m+n\tau)/\sqrt{5}$ with
  $\max(|m|,|n|)\leqslant 10^4$, where $I_{0}=I(0)$, for windows $W$
  of different lengths.  The dashed line corresponds to $k^4$ for
  length $\lvert W\rvert=\tau$ (top) and to $k^2$ for the other two
  cases.\label{fig:fibopeak}}
\end{figure}

Now, as $\ell\to\infty$, the same argument as above implies the
asymptotic behaviour
\[
      Z( \tau^{-\ell} k ) 
      \, \asymp \tau^{-4\ell}\, Z(k) \ts ,
\]
with the implied constants of type $\cO(1)$, and hence
$Z(k)=\cO(k^4)$, as $k\,\raisebox{2pt}{$\scriptscriptstyle
  \searrow$}\, 0$. This results means that, for inflation-invariant
projection sets, the distribution function $Z(k)$ of the diffraction
intensity vanishes like $k^4$ as
$k\,\raisebox{2pt}{$\scriptscriptstyle \searrow$}\, 0$, while, in the
generic case, we find a $k^2$-behaviour. This example illustrates that
the behaviour of the diffraction intensity near $0$ can pick up
non-trivial aspects of order in this system. This is illustrated for
some cases in Figure~\ref{fig:fibopeak}.

Our discussion above may appear quite special, in the sense that we
chose all scattering strengths to be equal. However, since we are only
interested in the scaling behaviour near the origin, this is in fact
no restriction, because the scaling law is unaffected by changing the
scattering strengths (as the length of the total window falls into
$\ZZ[\tau]$ if and only if the lengths of the sub-windows do).  This
simultaneously points to a strength and a weakness of this quantity as
a measure of order. On the one hand, the scaling behaviour can detect
and distinguish the order in the spatial arrangement of atoms
irrespective of the scattering strengths of the atoms; on the other
hand, it cannot provide any information on the distribution of
different scatterers. For the latter, the knowledge of the intensities
of the Bragg peaks is required.

Let us briefly comment on the scaling behaviour for other prominent
examples of aperiodic order discussed in \cite{BG19b}. For noble means
inflations, we observe the same $k^4$-scaling as for the Fibonacci
tiling. The period doubling sequence, which is limit periodic, shows
$k^2$-scaling, and a range of scaling exponents is accessible for
substitutions of more than two letters. For the Thue--Morse sequence,
which is the paradigm of an inflation structure with singular
continuous diffraction, we do not obtain a power law, but an
exponential scaling behaviour which decays faster than any power; see
also \cite{BGKS} for more on the scaling of the spectrum for this
system.  Finally, the Rudin--Shapiro sequence, which has absolutely
continuous spectrum, shows a linear scaling behaviour, due to the
constant density of its diffraction measure.

\section*{Acknowledgements}

It is our pleasure to thank Claudia Alfes-Neumann, Natalie Priebe
Frank, Neil Ma\~{n}ibo, Bernd Sing, Nicolae Strungaru and Venta
Terauds for valuable discussions, and two anonymous referees for
useful comments and suggestions. This work was supported by the German
Research Foundation (DFG), within the CRC 1283 at Bielefeld
University, and by EPSRC, through grant EP/S010335/1.

\begin{footnotesize}

\end{footnotesize}

\end{document}